\documentstyle[prl,aps,psfig,floats,amsfonts,amssymb]{revtex}
\begin{document}
 \twocolumn[\hsize\textwidth\columnwidth\hsize\csname  @twocolumnfalse\endcsname
\draft
\title{Magnetotransport through a strongly interacting quantum dot}
\author
{T. A. Costi}
\address
{Institut Laue--Langevin, 6 rue Jules Horowitz, B.P. 156, 38042 Grenoble Cedex
9, France
}
\maketitle
\begin{abstract}
We study the effect of a magnetic field on the conductance through a
strongly interacting quantum dot by using the finite temperature extension
of Wilson's numerical renormalization group method to dynamical quantities.
The quantum dot has one active level for transport and is modelled
by an Anderson impurity attached to left and right electron reservoirs.
Detailed predictions are made for the linear conductance and the 
spin-resolved conductance as a function of gate voltage, temperature and 
magnetic field strength. A strongly coupled quantum dot in a magnetic 
field acts as a spin filter which can be tuned by varying the gate 
voltage. The largest spin-filtering effect is found in the range of 
gate voltages corresponding to the mixed valence regime of the Anderson 
impurity model.
\end{abstract}
\vskip2pc]


There is currently much interest in understanding the equilibrium and 
non-equilibrium transport properties of nanoscale size quantum dots
\cite{goldhaber.98,cronenwett.98,schmid.98,simmel.99,wiel.00}. 
Due to the small size of these dots (ca. $100{\rm nm}$ in diameter), their 
level spacing, $\delta$, is less than an order of magnitude smaller
than the charging energy, $U$, for adding an electron to such a dot.
Provided the quantization of levels on the dot is not destroyed by
a large coupling to the reservoirs, single electron effects, such as
Coulomb blockade are observed. A more subtle effect of large charging
energies, however, is the creation of new states of many-body character 
at the Fermi level at sufficiently low temperatures
by the Kondo effect \cite{hewson.93}. This allows a quantum dot to
transmit perfectly at low enough temperatures, even when no single-particle
charge excitation of the dot is in resonance with the chemical potential 
of the reservoirs\cite{glazman.88,ng.88}. The effect requires a 
spin degenerate state on the dot,
which can be achieved, for example, by adjusting the dot levels $\varepsilon_i$
with a gate voltage so that the dot contains an odd number 
of electrons. The validity of this picture has been demonstrated by recent 
experimental work \cite{goldhaber.98,cronenwett.98,schmid.98,simmel.99}. 
For example, the predicted anomalous enhancement of the conductance with
decreasing temperature \cite{glazman.88,ng.88} has been observed for an 
odd number of electrons on the dot, and the signatures of the Kondo 
resonance, and its splitting in a magnetic field, have been seen in 
the differential conductance of quantum dots with an odd number of 
electrons (and its absence for an even number). Recently, the low 
temperature unitarity limit of perfect transmission, corresponding 
to a conductance of $2e^{2}/h$ for a single channel, has been measured 
for a range of gate voltages \cite{wiel.00}. As in \cite{goldhaber.98},
the results for the zero field conductance were found to be in very
good agreement with the theoretical calculations based on the
Anderson impurity model (AM) \cite{costi.94}. 

There has been much less work done on the transport properties of
strongly interacting quantum dots in a magnetic field, especially
theoretically. Some recent calculations on the Kondo model
in a field yielded the spectral densities and magnetoconductance 
of a quantum dot in the Kondo regime \cite{costi.00}. Quantum dots, 
however, can be tuned into other
interesting regimes, such as the mixed valent regime, by
adjusting the gate voltage on the dot \cite{goldhaber.98}. 
In order to describe the full range of behaviour observed in quantum dots,
the more general AM is required. 
In this paper we address theoretically the effects of a magnetic field 
on transport through a quantum dot in the various regimes of interest
by starting from an AM. We also make predictions 
for the spin-resolved magnetoconductance. Although spin-resolved 
measurements have so far not been carried out for quantum dots, this 
is likely to change given the current experimental interest in 
realizing spin-resolved currents through mesoscopic systems \cite{ohno.99}.

{\em Model---}
A quantum dot consisting of a
single correlated level $\varepsilon_{d}$ with Coulomb repulsion $U$ and 
coupled to left and right free electron reservoirs via energy independent
lead couplings $\Gamma_L$ and $\Gamma_{R}$ can be reduced to an AM in which a 
correlated level couples with strength 
$\Gamma=\Gamma_{L}+\Gamma_{R}$ to just one free electron reservoir
\cite{glazman.88}:
\begin{eqnarray}
{\cal H} &=& \sum_{\sigma}\varepsilon_{d}d_{\sigma}^{\dagger}d_{\sigma} 
+ Ud_{\uparrow}^{\dagger}d_{\uparrow}d_{\downarrow}^{\dagger}d_{\downarrow}
+ g\mu_{B}HS_{z}^{d}\nonumber\\
  &+& \sum_{k}\varepsilon_{k}c_{k,\sigma}^{\dagger}c_{k,\sigma}+
         \sum_{k,\sigma}v(c_{k,\sigma}^{\dagger}d_{\sigma} +
          d_{\sigma}^{\dagger}c_{k\sigma}),\label{eq:AM}
\end{eqnarray}
Here, $\varepsilon_{d}$ denotes the topmost occupied level of
the dot, which we take to be responsible for transport. All other 
energy levels of the dot are assumed unimportant for transport
and are neglected. The gate voltage, $V_g$, on the dot, 
is linearly related to the level position $\varepsilon_d$,
$-eV_{g}\sim \varepsilon_{d}$.
The first two terms in ${\cal H}$ 
represent the correlated level, the third term is
a magnetic field coupling only to the impurity spin
$S_{z}^{d}=\frac{1}{2}(d_{\uparrow}^{\dagger}d_{\uparrow}
-d_{\downarrow}^{\dagger}d_{\downarrow})$, the fourth term is the
free electron reservoir (a particular linear combination of the
original left and right reservoirs, the other combination, 
decouples from the impurity and is therefore not included here) 
and the last term is the hybridization of the impurity to the 
reservoir giving a lead coupling $\Gamma=\Gamma_{L}+\Gamma_{R}$. 
The current, $I=I(T,H,V)$, is given by \cite{hershfield.91}
\begin{eqnarray}
&&I  =
c\sum_{\sigma}\int_{-\infty}^{+\infty} 
A_{\sigma}(\omega,T,H,V)(f_{L}(\omega)-f_{R}(\omega))d\omega.\label{eq:current}
\end{eqnarray}
$A_{\sigma}(\omega,T,H,V)$ is the spectral density of the dot, which for
$V>0$ is a non-equilibrium quantity,
$f_{L,R}$ are the Fermi functions of the left and right leads, whose 
chemical potentials are $\mu_{L,R}=\pm |e|V/2$, and, $V$ is the transport 
voltage across the dot. The constant 
$c$ is given by 
$c=4e\Gamma_{L}\Gamma_{R}/\hbar(\Gamma_{L}+\Gamma_{R})$.
We assume, from here on, symmetric coupling 
to the leads, $\Gamma_L=\Gamma_R$. The linear magnetoconductance, 
$G(T,H)=\sum_{\sigma}G_{\sigma}(T,H)=\lim_{V\rightarrow 0}dI/dV$, is written
as a sum of spin-resolved magnetoconductances, $G_{\sigma}$, where
\begin{eqnarray}
G_{\sigma}(T,H) &=&\frac{e^2}{\hbar}\Gamma
\int_{-\infty}^{+\infty} d\omega\; A_{\sigma}(\omega,T,H)
\left(-\frac{\partial f(\omega)}{\partial\omega}\right).\label{eq:conductance}
\end{eqnarray}
$A_{\sigma}(\omega,T,H)=\lim_{V\rightarrow 0}A_{\sigma}(\omega,T,H,V)$ 
is an equilibrium spectral density and is expressed in terms of the 
local level Green function, ${\cal G}_{d,\sigma}$, by
\begin{eqnarray}
&&A_{\sigma}(\omega,T,H)
=-\frac{1}{\pi}{\rm Im}\;
{\cal G}_{d,\sigma}(\omega+i\epsilon,T,H),
\label{eq:spin-resolved-sd}
\end{eqnarray}
where ${\cal G}_{d,\sigma}= 1/(\omega-\varepsilon_{d}-g\mu_{B}H\sigma/2+
i\Gamma - \Sigma_{\sigma}(\omega))$
and
where $\Sigma_{\sigma}(\omega)$ is the correlation part of the self-energy.

{\em Method and calculations---}
We calculate $A_{\sigma}(\omega,T,H)$ by using
Wilson's NRG method \cite{wilson.75+kww.80} extended
to finite temperature dynamics\cite{costi.94,costi.99,bulla.00},
with recent refinements \cite{bulla.98,hofstetter.00} which
improve the high energy features, such as the
single-particle excitations at $\omega\sim \varepsilon_{d}$ and 
$\omega\sim\varepsilon_{d}+U$, of the AM. 
%
%
%
\begin{figure}[t]
\centerline{\psfig{figure=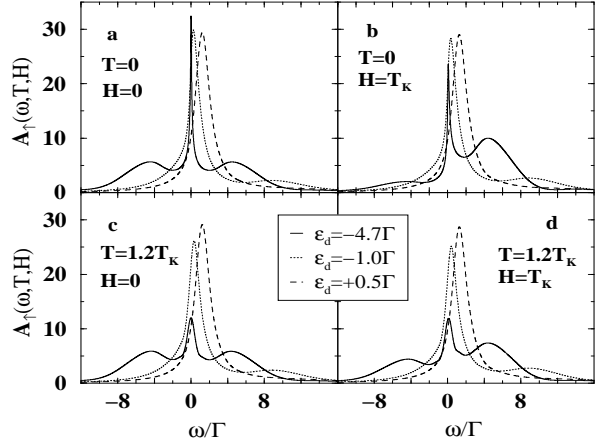,width=7.8cm,height=6.0cm,angle=0}}
\vspace{0.1cm}
\caption{
$A_{\uparrow}(\omega,T,H)$ in the Kondo ( 
$\varepsilon_{d}= -4.7\Gamma$), mixed valent 
($\varepsilon_{d}=-\Gamma$), and empty orbital ($\varepsilon_{d}=0.5\Gamma$) 
regimes at $T=0$ (a-b) and at $T=1.2T_{K}$ (c-d), where $T_{K}$ is defined
after Eq.\ (\ref{eq:ph-symmetry-sd}) 
}
\label{mt-fig1}
\end{figure}

\begin{figure}[t]
\centerline{\psfig{figure=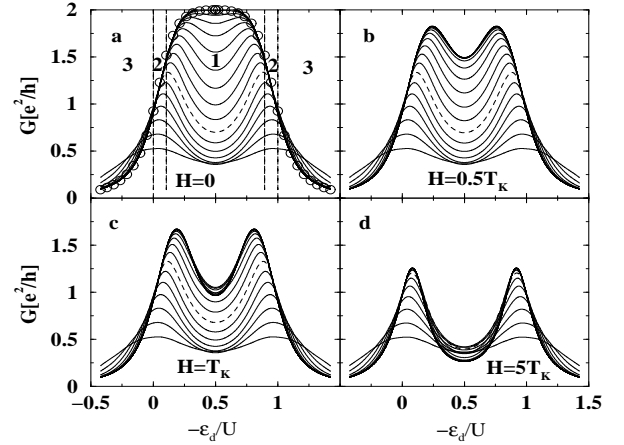,width=7.8cm,height=6.0cm,angle=0}}
\vspace{0.1cm}
\protect\caption{
Temperature and gate voltage dependence of 
the conductance $G(T,H)$for
fields (a) $H=0$, (b) $H=0.1T_{K}$, (c) $H=T_{K}$ and $H=5T_{K}$.
The symbols in (a) are the $T=0$ unitarity limit of the conductance
Eq.\ (\ref{eq:unitarity}). The temperatures decrease from bottom to top
and correspond to $T_N=T_0\Lambda^{-N}, N=0,1,2,\dots,\Lambda=1.5$ with the 
highest temperature being $T_0=1.32\Gamma$ (the dashed curves in (a-d) have
$T=1.2T_{K}$). The Kondo scale, $T_{K}$, is defined
after Eq.\ (\ref{eq:ph-symmetry-sd}). 
The regions marked $1,2$ and $3$ in (a)
and separated by vertical dashed lines correspond to the Kondo, 
mixed valent and empty orbital regimes respectively.
}
\label{mt-fig2}
\end{figure}

For the calculations we used $U/\pi\Gamma = 3$ and $\Gamma=0.01D$, 
with $D=1$ the half-bandwidth of the free electron reservoir, whose 
spectral density is taken to be independent of energy (we also set 
$|e|=g=\mu_{B}=k_{B}=1$ so that quantities like $H/T_{K}$, $V/\Gamma$ in the
figures should be interpreted as $g\mu_{B}H/k_{B}T_{K}$ and $|e|V/\Gamma$). 
This choice of parameters
suffices to account for all interesting effects of correlation.
Depending on the dot level position, there are three regimes of interest  
\cite{wilson.75+kww.80}:
(i) the Kondo regime, $-0.5U \leq \varepsilon_{d} \lesssim -\Gamma$, 
characterized by an exponentially small scale $T_{K}\sim(U\Gamma/2)^{1/2}
\exp(\pi\varepsilon_{d}(\varepsilon_{d}+U)/2\Gamma U)$, (ii) the mixed
valent regime $-\Gamma \lesssim \varepsilon_{d} \lesssim 0$ 
characterized by the charge fluctuation scale $\Gamma$ and, (iii),
the empty orbital regime $\varepsilon_{d}\gtrsim 0$ characterized by
a scale $\tilde{\varepsilon}_{d}\sim\varepsilon_{d}$, corresponding
to the renormalized level position. The region $\varepsilon_{d}\leq -0.5U$
is related to that for $\varepsilon_{d}\geq -0.5U$ by particle-hole 
symmetry:
${\cal H}(\varepsilon_{d},H)\leftrightarrow {\cal
  H}(-(\varepsilon_{d}+U),-H)$. As a consequence we have
\begin{eqnarray}
&&A_{\sigma}(\omega,T,H,\varepsilon_{d})
=A_{-\sigma}(-\omega,T,H,-(\varepsilon_{d}+U)).
\label{eq:ph-symmetry-sd}
\end{eqnarray}
We use this to calculate $G, G_{\uparrow}$ and $G_{\downarrow}$ at all
relevant $\varepsilon_d$. Throughout this paper, we define $T_K$ to be 
the HWHM of the $T=H=0$ Kondo resonance at the particle-hole symmetric
point $-\varepsilon_{d}/U=0.5$ (mid-valley point in Fig.\ \ref{mt-fig2}). 
It has the value $T_{K}\approx 0.001=\Gamma/10$.
%
%
%
%

{\em Spectral densities---}
Fig.\ \ref{mt-fig1}a-d summarizes the behaviour 
of the spectral density $A_{\sigma}(\omega,T,H)$ in zero and finite 
magnetic fields which will be useful for discussing the conductance 
results below (for full details see \cite{costi.94,costi.00,hofstetter.00}). 
The general trends in the Kondo regime at $H=0$ are two almost temperature 
independent resonances at $\omega\approx\varepsilon_{d}$ and 
$\omega\approx\varepsilon_{d}+U$ and a strongly temperature 
dependent Kondo resonance at $\omega\approx 0$ (cf. Fig.\ \ref{mt-fig1}a,c). 
Fields comparable to
$T_{K}$ result in a large shift of spectral weight in $A_{\uparrow}$ from
the excitation at $\varepsilon_{d}$ to the one at $\varepsilon_{d}+U$
(and an opposite shift in $A_{\downarrow}$). 
The Kondo resonance in $A_{\uparrow}$ ($A_{\downarrow}$) shifts upwards 
(downwards) by an amount $\lesssim H$ and is reduced in height \cite{costi.00} 
(cf. Fig.\ \ref{mt-fig1}a,b). In the mixed valent regime, the Kondo resonance
has merged with the excitation at $\varepsilon_{d}$ to form a renormalized
resonance of width $\Gamma$ lying close to but above the Fermi level at
$\tilde{\varepsilon}_{d}< \Gamma/2$. The 
excitation at $\varepsilon_{d}+U$ has little weight. The resonance at
$\tilde{\varepsilon}_{d}\approx\varepsilon_{d}$ in the empty orbital 
regime is almost independent of temperature and field on scales 
comparable to $\Gamma$ and the one at $\varepsilon_{d}+U$ is no longer 
discernable.

{\em Gate voltage dependence of the conductance---}
The gate voltage dependence of the linear
conductance is shown in Fig.\ \ref{mt-fig2}a-d for several temperatures
and magnetic field strengths. Consider first the zero field case.
The charging energy, $U$, corresponds to the separation between the 
Coulomb blockade peaks at high temperature (these arise when $\varepsilon_d$
or $\varepsilon_{d}+U$ coincide with the Fermi level of the leads).
On lowering the temperature these peaks move together as a result
of the development of the Kondo resonance in region 1, as observed 
experimentally \cite{goldhaber.98}. The conductance in region 1 
continues to increase with decreasing temperature, eventually 
reaching the unitarity limit of $2e^{2}/h$, which has recently been seen
in experiments \cite{wiel.00}. As
a check on our $T\rightarrow 0$ results for all $\varepsilon_{d}$ we 
have shown the $T=0$ unitarity curve \cite{glazman.88,ng.88} (which
follows from \cite{langreth.66})
\begin{eqnarray}
G(T=H=0)&=&\frac{2e^2}{h}\sin^{2}(\pi n_d/2),\label{eq:unitarity}
\end{eqnarray}
with the dot level occupancy, $n_{d}$, deduced from the spectral densities. 
As in the experiments \cite{goldhaber.98}, the enhancement of the 
conductance with decreasing temperature in the
Kondo and mixed valent regimes (regions 1 and 2 in Fig.\ \ref{mt-fig2})
is in marked contrast to the suppression of the conductance with decreasing
temperature in the empty orbital case (region 3).
%
%
%
%
\begin{figure}[t]
\centerline{\psfig{figure=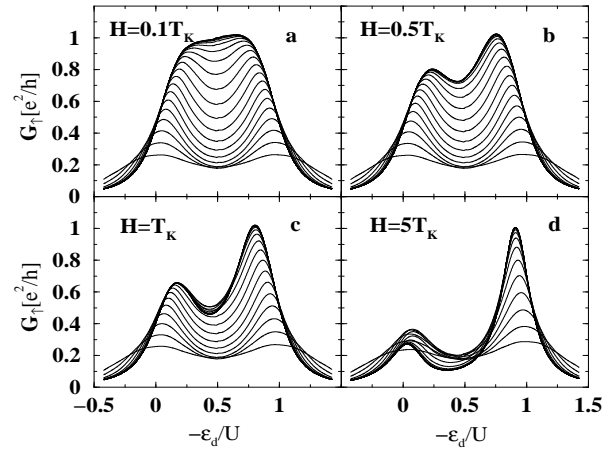,width=7.8cm,height=6.0cm,angle=0}}
\vspace{0.1cm}
\caption{
Dependence of the spin-resolved conductance $G_{\uparrow}(T,H)$ on 
gate voltage for several magnetic
fields (a-d), and for the same $T_K$ and temperatures as in 
Fig.\ \ref{mt-fig2}. 
$G_{\downarrow}$ is the mirror reflection of $G_{\uparrow}$
about $\varepsilon_{d}/U=-0.5$ as described in the text. 
}
\label{mt-fig3}
\end{figure}

On applying a small magnetic field we see a drastic change in the
total conductance in the Kondo regime. A magnetic field suppresses the 
Kondo effect, tending to split the total spectral density at the Fermi 
level, and thereby decreases
the conductance in the Kondo valley (region 1). A sizeable effect is also seen
in the mixed valent regime, but very little change is observed in the
empty orbital case. The effects of a magnetic field become even more 
apparent in spin-resolved quantities such as $G_{\uparrow}$ in 
Fig.\ \ref{mt-fig3}
(from Eq.\ (\ref{eq:ph-symmetry-sd}) $G_{\downarrow}$, as a function 
of $\varepsilon_{d}$, is the mirror reflection of $G_{\uparrow}$ about 
$\varepsilon_{d}=-0.5U$). Again, the main effect of a field is to
change dramatically the conductance in the Kondo regime (region 1 of
Fig.\ \ref{mt-fig2}a) with some changes also in the mixed valent regime.
The low $T$ asymmetry in the conductance $G_{\uparrow}$ about 
$\varepsilon_d=-0.5U$ at finite fields reflects the fact that the
$H=0$ Kondo resonance lies just above the Fermi level for 
$\varepsilon_{d}>-0.5U$
but just below the Fermi level for $\varepsilon_{d}<-0.5U$. Consequently,
applying a field $+H$, which moves the Kondo resonance in $A_{\uparrow}$ 
upwards, leads to a suppression of $G_{\uparrow}$ for the former
case and an enhancement for the latter. A quantum dot in a field is
seen to act as a spin-filter as discussed in \cite{recher.00} for
dots weakly coupled to leads ($G_{\sigma}\ll e^2/h$). We note that
the effect is largest in the mixed valent regime (see Fig.\ \ref{mt-fig4}
below).
%
%
%
%
\begin{figure}[t]
\centerline{\psfig{figure=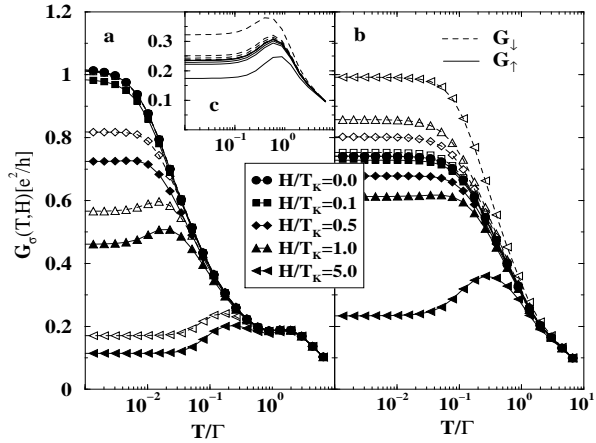,width=7.8cm,height=6.0cm,angle=0}}
\vspace{0.1cm}
\protect\caption{
Temperature dependence of $G_{\uparrow}$ (filled symbols, solid lines) and
$G_{\downarrow}$ (open symbols, dashed lines) at several
values of the magnetic field for three 
characteristic gate voltages corresponding to (a) the Kondo regime 
($\varepsilon_{d}=-4.7\Gamma$), (b), the mixed valent regime 
($\varepsilon_{d}=-\Gamma$), and, (c), the empty orbital regime 
($\varepsilon_{d}=+\Gamma$, symbols omitted for clarity). 
Dashed (solid) curves are for 
$G_{\downarrow}$ ($G_{\uparrow}$). The Kondo scale, $T_{K}$, is that
defined in Fig.\ \ref{mt-fig2}.
}
\label{mt-fig4}
\end{figure}

{\em Temperature dependence of the conductance---}
Fig.\ \ref{mt-fig4} shows the temperature dependence of the
spin-resolved conductance at several magnetic fields and for 
gate voltages corresponding to Kondo (Fig.\ \ref{mt-fig4}a),
mixed valent (Fig.\ \ref{mt-fig4}b) and empty orbital regimes 
(inset to Fig.\ \ref{mt-fig4}a). The results for the Kondo regime
are similar to those calculated directly from the Kondo model \cite{costi.00},
except that charge fluctuations, absent in the former, give rise, 
in the AM, to a small peak at $T\gtrsim\Gamma$ in $G_{\sigma}$. 
At low temperatures, $T < T_{K}$, the two spin components of the Kondo
resonance move away from the Fermi level and decrease in height
with increasing field \cite{costi.00}. This leads to the observed strong 
suppression of {\em both} 
$G_{\uparrow}(T< T_{K})$ and $G_{\downarrow}(T< T_{K})$ 
and to the appearance for $H\gtrsim 0.5T_{K}$ of a noticeable peak 
at $T\approx H$ and would be experimentally observable in transport
measurements of $G(T,H)$.  
The behaviour in the mixed valent (Fig.\ \ref{mt-fig4}b) and
empty orbital (inset to Fig.\ \ref{mt-fig4}a) regimes is
different. $G_{\uparrow}$ is suppressed whereas $G_{\downarrow}$ is
enhanced by increasing the field. This is understood
by noting that for these cases, the resonance in the spectral
density at $H=0$ lies above the Fermi level (see Fig.\ \ref{mt-fig1}), 
and application of a field has little effect on the height of the
spin components, $A_{\sigma}$,  of the spectral density
(in contrast to the Kondo regime). The main
effect is to shift these components in opposite directions,
resulting in $A_{\downarrow}(\omega=0,T=0,H)$ increasing 
and $A_{\uparrow}(\omega=0,T=0,H)$ decreasing
with increasing field, which thereby gives rise to the
opposite trends in $G_{\uparrow}$ and $G_{\downarrow}$. 
The main difference between the mixed valent and empty orbital cases, is 
the stronger field dependence of $G_{\sigma}$ 
for the former.

In summary we have studied magnetotransport through a strongly
interacting quantum dot with one active level for transport. 
Quantitative agreement with the $H=0$ conductance as a function
of temperature has already been demonstrated for quantum dots in 
heterostructures \cite{goldhaber.98,wiel.00} and more recently also 
for quantum dots defined in carbon nanotubes \cite{nygard.00}. Our
predictions for the conductance in a magnetic field can be directly
tested with measurements on quantum dots with a field 
parallel to the plane of the quantum dot so that
orbital effects can be neglected. Such 
experiments are currenlt being planned \cite{private-communication}.
There is also much interest in realizing spin polarized currents 
through mesoscopic devices, such as quantum dots, so our 
results for spin-resolved conductances in a magnetic field could
eventually be tested against experiment. Our results show
that a strongly coupled quantum dot ($G_{\sigma}\sim e^{2}/h$) 
in a magnetic field acts as a spin-filter as found for weakly coupled
($G_{\sigma}\ll e^{2}/h$) quantum dots \cite{recher.00}. 
The largest spin-filtering effect is in the mixed valent regime.

Useful discussions with J. von Delft, S. De Franceschi, D. Goldhaber-Gordon,
E. Kats, P. Kr\"{u}ger, D. Loss, Ph. Nozi\`eres, F. Pistolesi and
E. Sukhorukov are gratefully acknowledged.


\end{document}